\documentclass[preprints,article,accept,pdftex,moreauthors,galaxies]{Definitions/mdpi} 
\usepackage{aas_macros}
\usepackage{graphbox}
\firstpage{1} 
\pubvolume{1}
\issuenum{1}
\articlenumber{0}
\pubyear{2022}
\copyrightyear{2022}
\datereceived{} 
\dateaccepted{} 
\datepublished{} 

\hreflink{https://doi.org/} 

\Title{Tracing the hot spot motion using the next generation Event Horizon Telescope (ngEHT)}

\TitleCitation{Emami, R.; Tiede, P.; Doeleman, S.S.; Roelofs, F.; Wielgus, M.; Blackburn, L.; Liska, M.;
Chatterjee, C.; Fuentes, A.; Broderick, A.; 
Hernquist, L.; Alcock, C.; 
Narayan, R.;  Smith, R.; Tremblay, G.; Ricarte, A.; Sun, h.; Anantua, R.; Kovalev, Y.Y.; Natarajan, P.; Vogelsberger, M.; hot spot motion extraction with ngEHT, }

\Author{Razieh Emami $^{1}$\orcidA{}, Paul Tiede $^{1,2}$, Sheperd S. Doeleman $^{1,2}$, Freek Roelofs$^{1,2}$, Maciek Wielgus$^{3}$, Lindy Blackburn$^{1,2}$, Matthew Liska$^{1}$, Koushik Chatterjee$^{1,2}$, 
Bart Ripperda $^{4,5}$, 
Antonio Fuentes$^{6}$, Avery Broderick $^{7,8}$, Lars Hernquist$^{1}$, Charles Alcock$^{1}$, Ramesh Narayan $^{1,2}$, Randall Smith$^{1}$, Grant Tremblay$^{1}$,  Angelo Ricarte $^{1,2}$, He Sun $^{9}$, Richard Anantua$^{10}$, Yuri Y. Kovalev$^{3,11,12}$, Priyamvada Natarajan $^{2,13,14}$, Mark Vogelsberger $^{15}$}

\address{$^{1}$ \quad  Center for Astrophysics $\vert$ Harvard \& Smithsonian, 60 Garden Street, Cambridge, MA 02138, USA \\
$^{2}$ \quad Black Hole Initiative at Harvard University, 20 Garden Street, Cambridge, MA 02138, USA \\
$^{3}$ \quad Max-Planck-Institut f\"ur Radioastronomie, Auf dem H\"ugel 69, D-53121 Bonn, Germany \\
$^{4}$ \quad School of Natural Sciences, Institute for Advanced Study, 1 Einstein Drive, Princeton, NJ 08540, USA \\
$^{5}$ \quad NASA Hubble Fellowship Program, Einstein Fellow \\
$^{6}$ \quad  Instituto de Astrofísica de Andalucía-CSIC, Glorieta de la Astronomía s/n, E-18008 Granada, Spain \\
$^{7}$ \quad Perimeter Institute for Theoretical Physics, 31 Caroline Street North, Waterloo, ON, N2L 2Y5, Canada \\
$^{8}$ \quad  Department of Physics and Astronomy, University of Waterloo, 200 University Avenue West, Waterloo, ON, N2L 3G1, Canada \\
$^{9}$ \quad National Biomedical Imaging Center, College of Future Technology, Peking University, China \\
$^{10}$ \quad  Department of Physics $\&$ Astronomy, The University of Texas at San Antonio, One UTSA Circle, San Antonio, TX 78249, USA \\ 
$^{11}$ \quad Lebedev Physical Institute of the Russian Academy of Sciences, Leninsky prospekt 53, 119991 Moscow, Russia\\
$^{12}$ \quad Moscow Institute of Physics and Technology, Institutsky per. 9, Dolgoprudny 141700, Russia\\
$^{13}$ \quad Department of Astronomy, Yale University, New Haven, CT 06511, USA\\
$^{14}$ \quad Department of Physics, Yale University, New Haven, CT 06520, USA\\
$^{15}$ Department of Physics, Kavli Institute for Astrophysics and Space Research, Massachusetts Institute of Technology, Cambridge, MA 02139, USA
}

\corres{razieh.emami$_{-}$meibody@cfa.harvard.edu}

\abstract{We propose to trace the dynamical motion of a shearing hot spot near the SgrA* source through a dynamical image reconstruction algorithm, StarWarps. Such a hot spot may form as the exhaust of magnetic reconnection in a current sheet near the black hole horizon. A hot spot that is ejected from the current sheet into an orbit in the accretion disk may shear and diffuse due to instabilities at its boundary during its orbit, resulting in a distinct signature. We subdivide the motion to two distinct phases; the first phase refers to the appearance of the hot spot modelled as a bright blob, followed by a subsequent shearing phase simulated as a stretched ellipse. We employ different observational arrays, including EHT(2017,2022) and the next generation event horizon telescope (ngEHTp1, ngEHT) arrays, in which few new additional sites are added to the observational array. We make dynamical image reconstructions for each of these arrays. Subsequently, we infer the hot spot phase in the first phase followed by the axes ratio and the ellipse area in the second phase. We focus on the direct observability of the orbiting hot spot in the sub-mm wavelength. Our analysis demonstrates that newly added dishes may easily trace the first phase as well as part of the second phase, before the flux is reduced substantially. The algorithm used in this work can be extended to any other types of the dynamical motion. Consequently, we conclude that the ngEHT is a key to directly observe the dynamical motions near variable sources, such as SgrA*. }

\keyword{hot spot, Dynamical image reconstruction, SgrA*, Time-variability, EHT, ngEHT, StarWarps}

\begin{document}
\thispagestyle{empty}

\section{Modelling flares in Sgr~A* with hot spots}
The recent resolved images of Sagittarius A* (SgrA*) by the Event Horizon Telescope (EHT) \citep{2022ApJ...930L..12A, 2022ApJ...930L..13A, 2022ApJ...930L..14A, 2022ApJ...930L..15A,2022ApJ...930L..16A,2022ApJ...930L..17A,2022ApJ...930L..18F, 2022ApJ...930L..19W, 2022ApJ...930L..20G, 2022ApJ...930L..21B} revealed rapid structural variability of the resolved super massive black hole (SMBH) source at the galactic center \cite{2019BAAS...51g.256D,2019BAAS...51g.235J}. These findings complement the reported variability of this compact source across electromagnetic spectrum \citep{2021ApJ...917...73W}, in the mm/sub-mm \citep{2008Natur.455...78D,2008JPhCS.131a2055D,2008arXiv0807.2427F,Marrone2008,2022ApJ...930L..19W,2009ApJ...695...59D,2014IAUS..303..288A,2014ApJ...794..150J,2014ApJ...795..134F}, in near-infrared (NIR) \cite{2003Natur.425..934G, 2006A&A...455....1E, 2019ApJ...882L..27D} and X-ray \citep{2001Natur.413...45B,  2003A&A...407L..17P, 2022MNRAS.510.2851A, 2019ApJ...886...96H,2011ApJ...726...54K,2017MNRAS.472.4422K}. In particular, during the flare events, flux density observed in NIR and X-ray increases by 1-2 orders of magnitude, which roughly aligns with theoretical expectations, e.g., \citep{2020MNRAS.497.4999D}. The flares seem to originate from a compact region near the innermost stable circular orbit (ISCO) \citep{gravity_loops_2018, Wielgus2022b}. 

In particular, \citep{Wielgus2022b} recently reported an orbiting hot spot detection in the unresolved light curve data at the EHT observing frequency, following an X-ray flare. 

Theoretically, there have been various explorations trying to model these flares (hot spot) from general relativistic magnetohydrodynamical (GRMHD) or through some semi-analytic models. In the former case, magnetic reconnection and the flux eruption \citep{2014ARA&A..52..529Y, 2020MNRAS.497.4999D} are good candidates to produce such flares, in a form of a hot spot region, orbiting around the SMBH, arisen from the local energy injection accelerating the electrons \citep{2004ApJS..153..205D,2005MNRAS.363..353B,2006MNRAS.367..905B,2020ApJ...892..132T} within the accretion disk. While in the later case, the hot spot may be embedded within a geometrically thick, hot and optically-thin radiatively-inefficient accretion flow (RIAF; \citep{1982Natur.295...17R, 1994ApJ...428L..13N, Narayan+1995, 2019ApJS..243...26P}) expected to be characteristic of low-luminosity SMBHs such as Sgr~A*.

\begin{figure*}[th!]
\centering
\includegraphics[width=\textwidth]{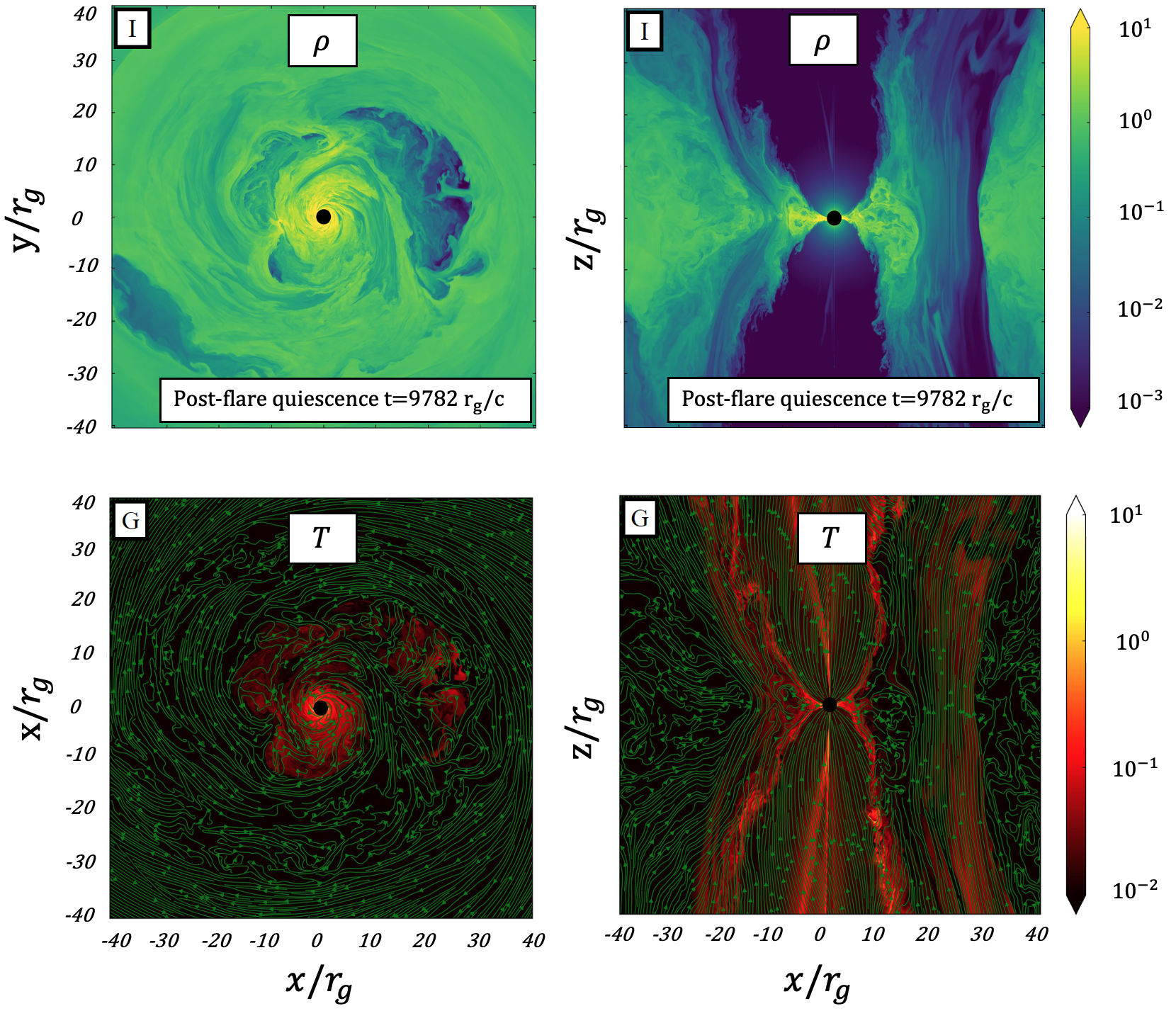}
\caption{The formation of plasmoids, in HAMR simulation \cite{2022ApJ...924L..32R}, from tearing instability in a very thin equatorial current sheet formed as a consequence of squeezing the accretion under the conversion of horizontal field lines to the vertical ones. In this scenario the field conversion is owing to the reconnection and results in an exhausted low density hot spot confined by the vertical field.  }
\label{fig:Plasmoid-hotspot}
\end{figure*}

\section{Dynamical formation of hot spot in the simulations}
Formation of hot spots has been reported in general relativistic magnetohydrodynamics (GRMHD) simulations. In these simulations, as the gas near the black hole becomes more magnetized reaching the MAD state, horizontal fields squeeze the accretion flow, thereby forming a thin equatorial current sheet \citep{2022ApJ...924L..32R}. This current sheet is potentially unstable to tearing instabilities and the formation of plasmoids via reconnection. Plasmoids are relativistically hot blobs of plasma that are surrounded by more magnetized gas.

In a scenario proposed by \citep{2022ApJ...924L..32R}, an equatorial reconnection layer transforms horizontal field at the jet base, into vertical field that is injected into the accretion disk. The flux tube of vertical field is filled with non-thermal leptons, originating from the jet's magnetized plasma, and accelerated by the reconnection. The resulting low-density hot spot contained by vertical magnetic field is pushed into orbit around the black hole and conjectured to power NIR emission, trailing a large X-ray flare. Figure \ref{fig:Plasmoid-hotspot} presents the dynamical formation of the hot spot filled with low-density plasma contained by vertical field from a HAMR simulation \citep{2022ApJ...924L..32R}.

Large plasmoids, formed due to mergers of smaller plasmoid in reconnection layers have also been conjectured as a model for orbiting hot spots. The growth and propagation of plasmoids is still an ongoing area of research especially in full 3D GRMHD. Because of the potential of these plasmoids to carry non-thermal electrons (as magnetic reconnection can drive particle acceleration), a number of works have tried to model plasmoid evolution as spherical or shearing hot spots around black holes \citep{hotspot1,2006JPhCS..54..448B, 2006JPhCS..54..443M,2007IAUS..238..407M,2008JPhCS.131a2008Z,2011ApJ...735..110B,2015MNRAS.454.3283Y,2020ApJ...892..132T}. 

The main difference between the vertical flux tube scenario and an individual large plasmoid as a hot spot model is twofold: a plasmoid consists of dominantly helical field and is shown to mainly orbit along the jet sheath \citep{Nathanail2020,Ripperda2020}; whereas a large flux tube formed as reconnection exhaust consists of vertical field and orbits in the accretion disk. Recent observations of orbiting hot spots seem to suggest a dominant vertical field component \cite{gravity_loops_2018,Wielgus2022b} associated with the motion, that implies that a vertical field flux tube may be more realistic as the source of emission, instead of an individual large plasmoid. On the other hand, in a different scenario an apparent hot spot observed at mm wavelengths could correspond simply to a local density maximum, possibly randomly originating in the turbulent accretion flow or related to an infalling clump of matter \citep{Moriyama2019}.
\begin{figure*}[th!]
\centering
\includegraphics[width=\textwidth]{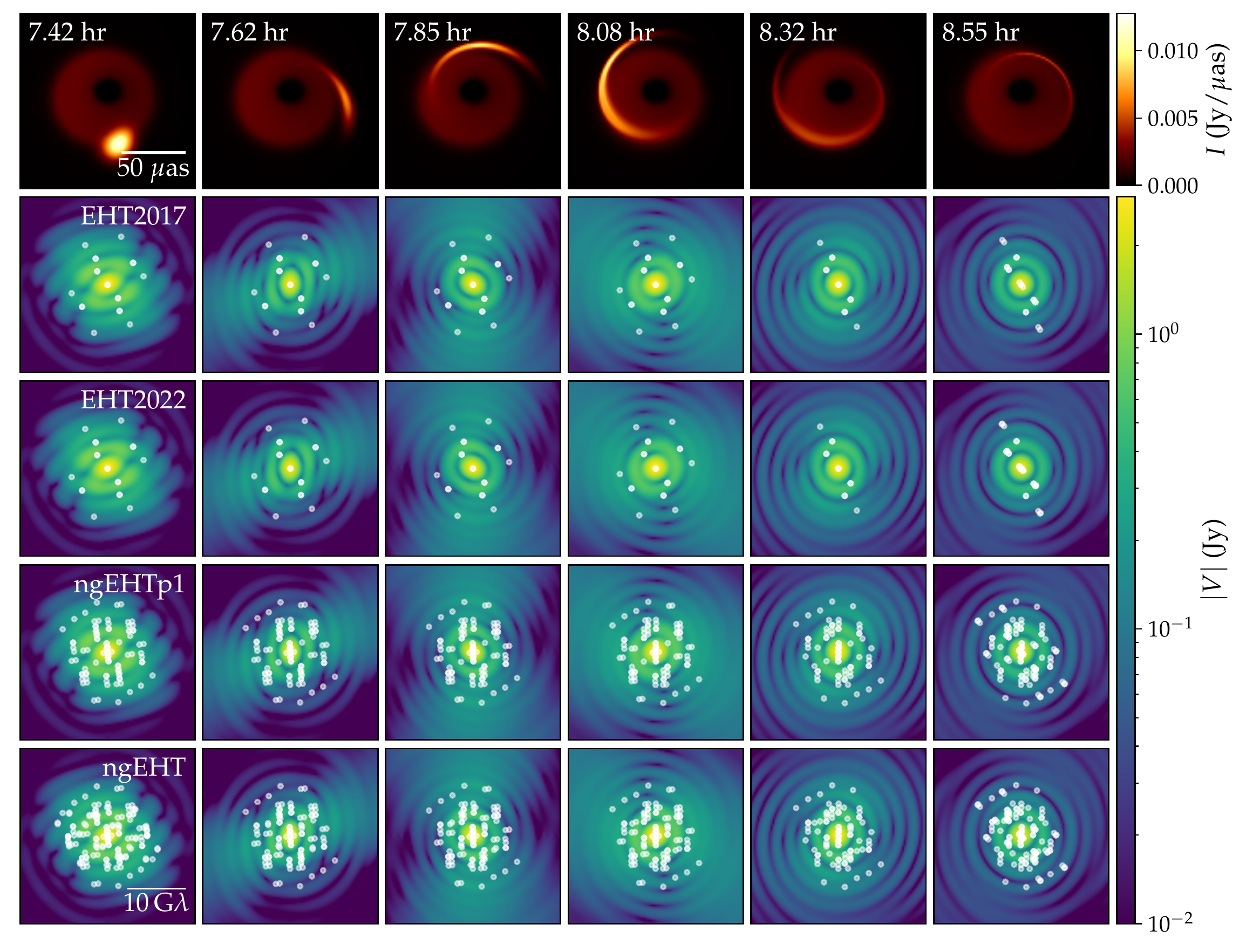}
\caption{Simulation used for tracing hot spot motion. The top row shows 6 frames from the simulation equally spaced from 7.42 UT to 8.55 UT. The following 4 rows shows the visibility amplitudes of the movie frames for different array configurations (white dots).  }
\label{fig:original-hotspot}
\end{figure*}

\section{Semi-analytic simulation of a shearing hot spot}
There have been a variety of different hot spot models. The original studies \cite{2005MNRAS.363..353B, 2006MNRAS.367..905B} only focused on the coherent motion of a spherical Gaussian hot spot. \cite{2009A&A...500..935E} extended this model by adding the adiabatic expansion and \cite{2010A&A...510A...3Z} considered a 2D shearing hot spot, ignoring the radiative transfer effects. More recently, \cite{2020ApJ...892..132T} extended this model further and included both the shearing and the expansion of a 3D hot spot, additionally incorporating some radiative transfer effects, while \cite{Vos2022} focused on employing a full polarized radiative transfer. The model assumes that the hot spot satisfies the continuity equation, and travels along a prescribed velocity field $u^\mu$. If we further assume that $u^\mu$ has limited vertical motion and is axisymmetric, the evolution of the hot spot electron number density $n_e$ can be written as:
\begin{equation}
\label{hotspot-model}
n_e(\tau, x^{\mu}) = n_{e0}(y^{\mu}) \sqrt{\frac{-g(y^{\mu})}{-g(x^{\mu})}} \frac{u^{r}(y^{\mu})}{u^{r}(x^{\mu})}.
\end{equation}
where $n_{e0}(y^{\mu})$ refers to the initial proper density of hot spot, $y^{\mu}$ describes its initial location and $x^{\mu}$ refers to its subsequent position. Note that $y^\mu = \phi_{-\tau}(x^\mu)$, where $\phi_\tau$ is the velocity field flow found by integrating $\dot x^\mu = u^\mu$ for $\tau$ units of proper time.  
As the hot spot model provides an efficient framework in tracking the flares, it is really intriguing to consider its direct observability in the sub-mm array.

This hot spot model provides a clean framework to measure how well different VLBI arrays can measure the plasma dynamics near the black hole.  

Figure \ref{fig:original-hotspot} presents the appearance of the shearing hot spot at few different times in the image space (top row) and in the visibility space using EHT2017, EHT2022,  phase I of ngEHT (ngEHTp1) and the full array of  ngEHT (ngEHT), respectively.

Motivated by these aforementioned  theoretical and observational studies, in this paper, we employ to use this extended hot spot model and propose to directly observe this through dynamically reconstructing the hot spot motion using the StarWarps algorithm \cite{2017arXiv171101357B} (see below for more details). 

We use few different observational arrays focusing on the prospects of direct observability of the orbiting hot spot in the sub-mm wavelength. This includes both of the current EHT coverage (EHT2017, EHT2022) as well as the next generation of the event horizon telescope (ngEHT) arrays (ngEHTp1, ngEHT) with new multiply added sites around the globe. 

We conclude that the hot spot motion can be traced by the next generation of the Event Horizon Telescope. Finally, while we only focus on tracing the hot spot motion in this work, the methods and underlying analysis can also be generalized to almost any types of the dynamical motions.

\section{Creating synthetic data for EHT/ngEHT} \label{synthetic-data}
To make the synthetic data for the dynamical image reconstruction, we made use of the ehtim package \citep{2018AAS...23134721P, 2018zndo...1173414C, 2022zndo...6519440C}. Our array contains 4 different subsets including the EHT(2017), EHT(2022), ngEHTp1 and ngEHT. Representative April weather is used to simulate station performance, along with random (uncalibrated) absolute atmospheric phase and $\sim$10\% amplitude gain systematic error. \autoref{tab:arrayconfiguration} contains a list of stations used for each array configuration.

\begin{figure*}[th!]
\centering
\includegraphics[width=0.4\textwidth]{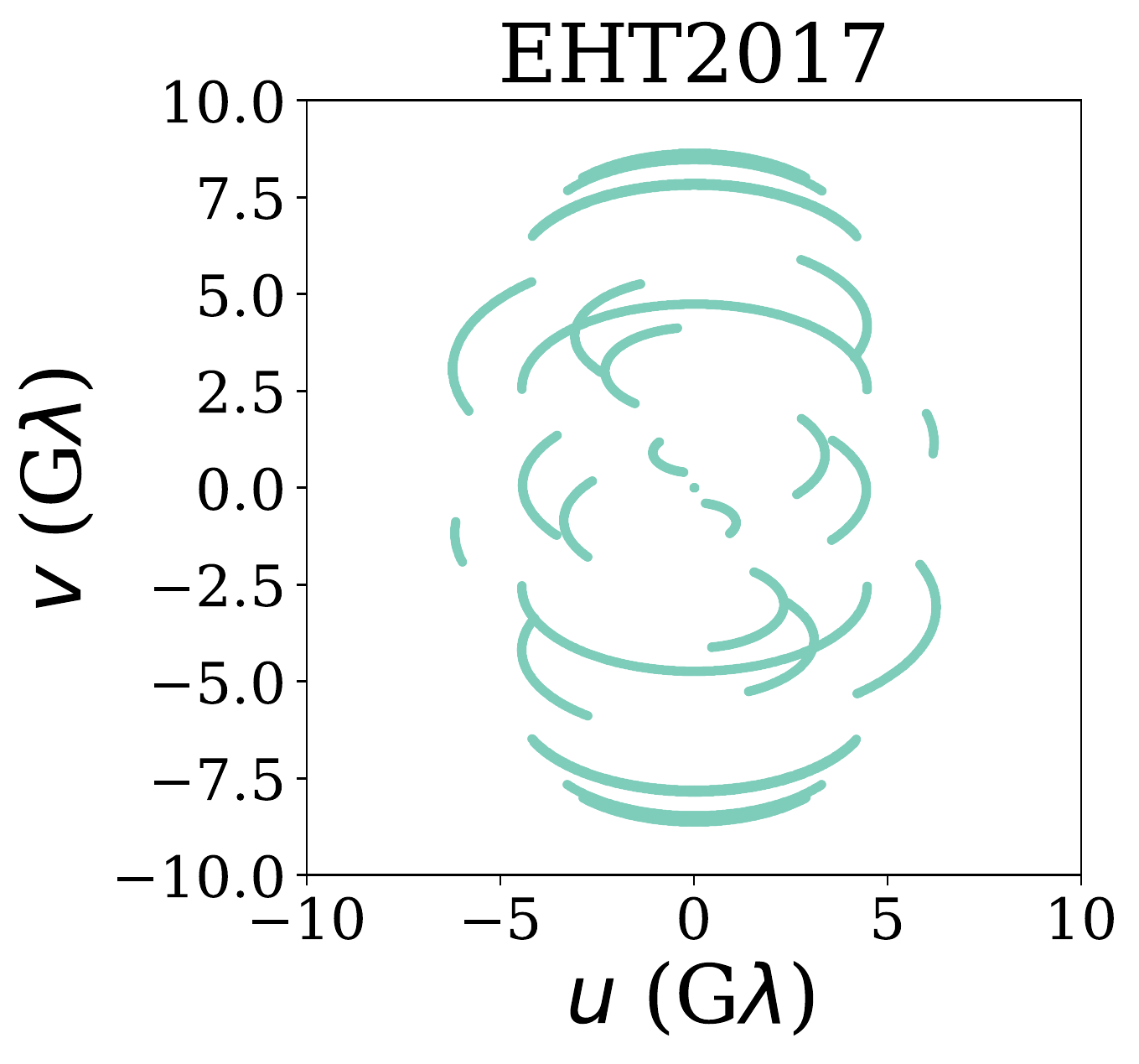}
\includegraphics[width=0.4\textwidth]{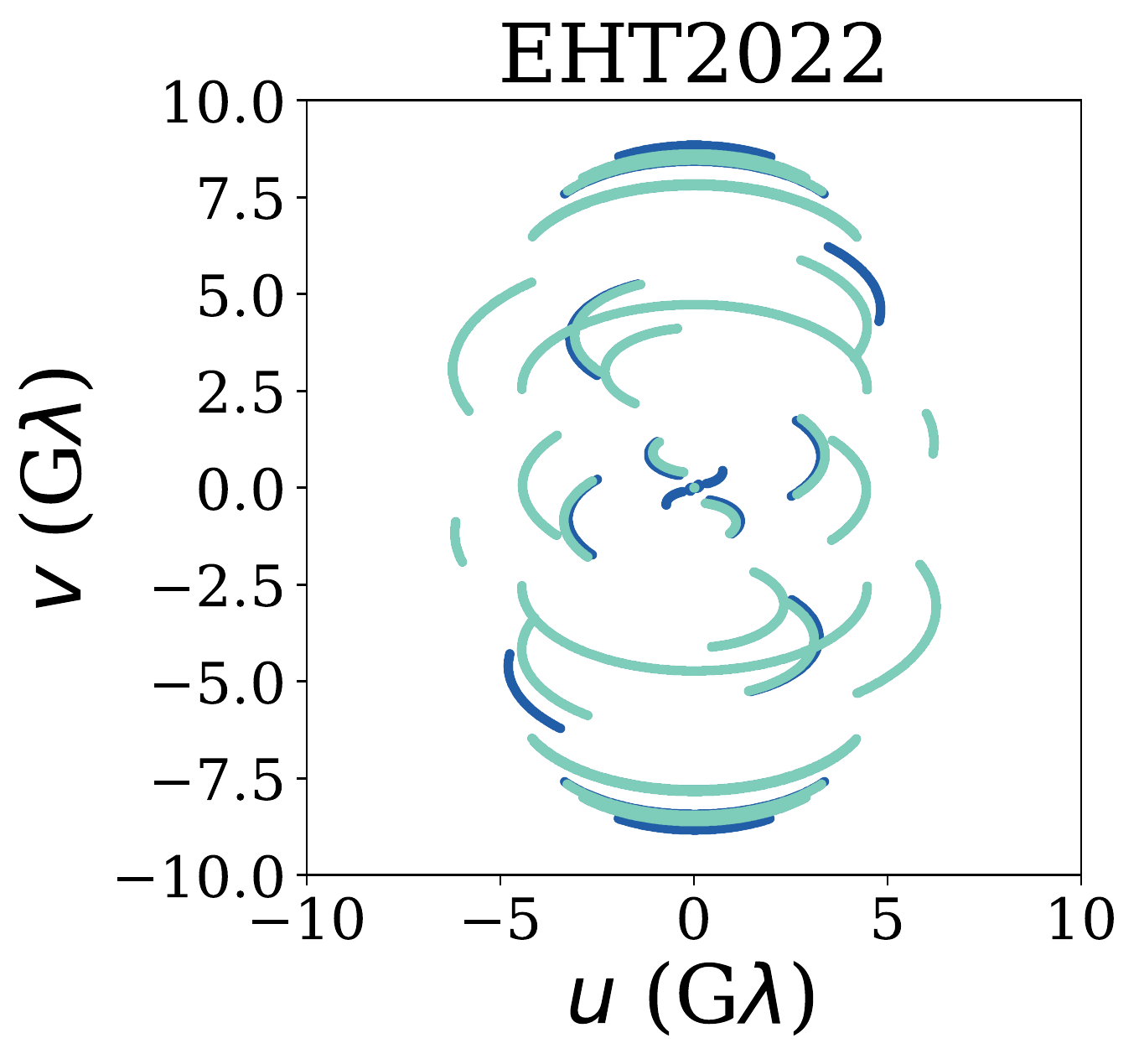}
\includegraphics[width=0.4\textwidth]{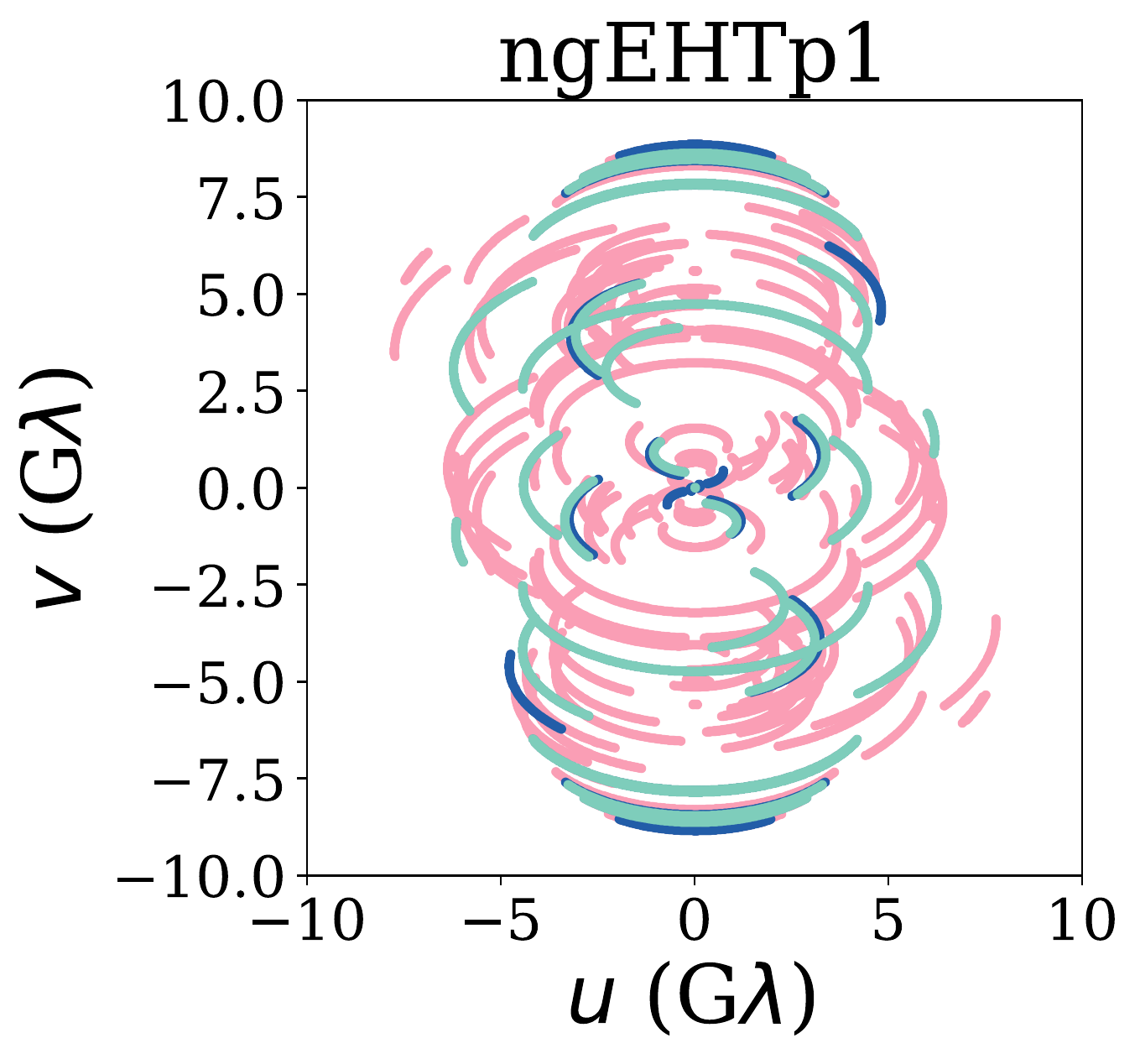}
\includegraphics[width=0.4\textwidth]{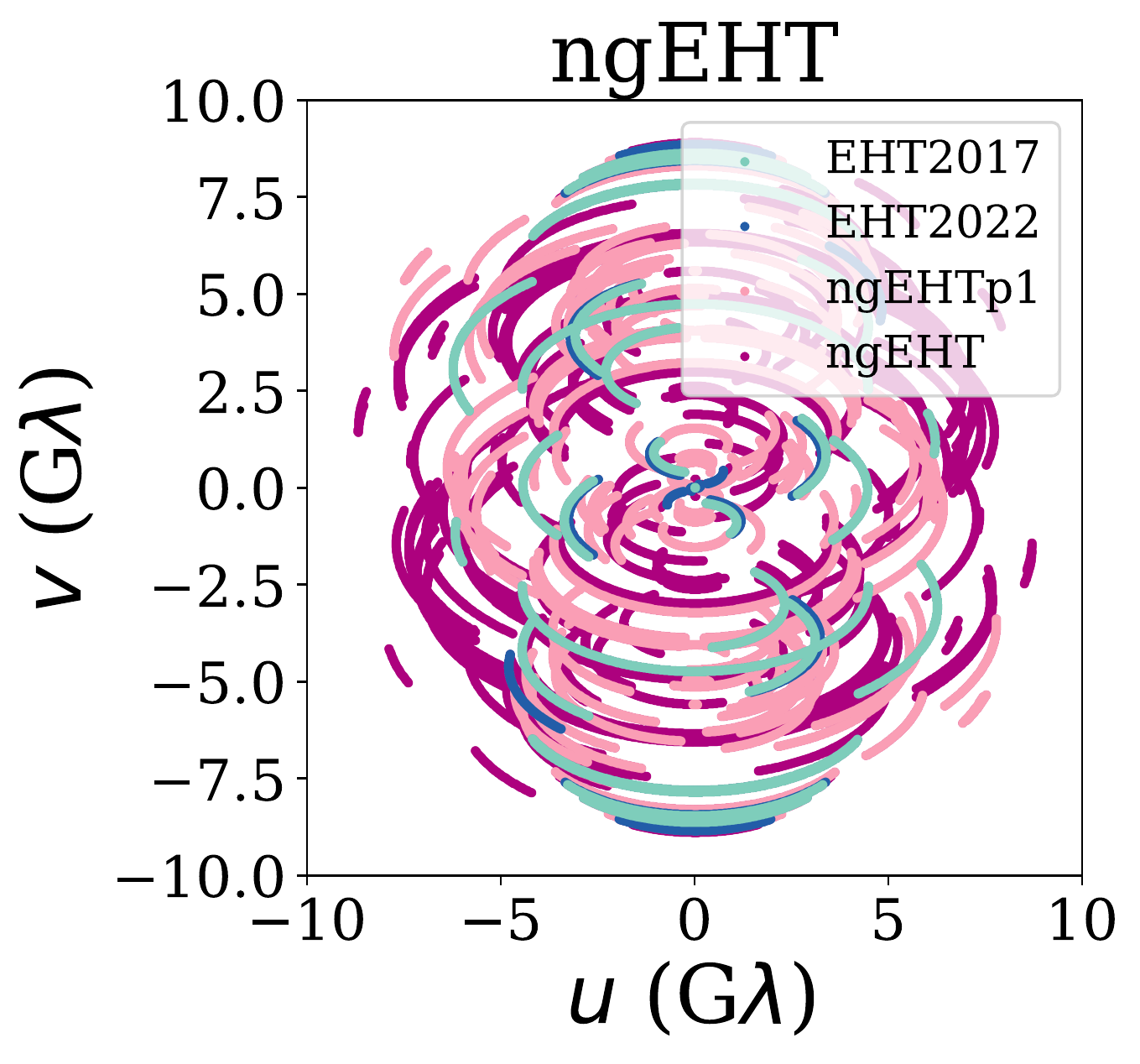}
\caption{The uv-coverage for different observational arrays. The top row presents the uv-coverage for EHT(2017) (left panel) and EHT(2022) (right panel), while the bottom row shows the coverage for ngEHTp1 (left) and the full ngEHT (right) arrays, respectively. It is clearly seen that adding the new sites/dishes significantly improves the uv-coverage in the observational array. As it is shown below, the improved coverage is significantly useful in tracing the orbital motion of hot spot.}
\label{fig:uv-coverage}
\end{figure*}

\begin{table}
\begin{tabular}{c|ccccccccc}
Array & \multicolumn{8}{c}{Sites used for simulated data} & \\
\hline
  EHT(2017) & ALMA & APEX & SMA & JCMT & SMT & LMT & PV & SPT \\
  EHT(2022) & EHT(2017)+ & KP & GLT & NOEMA \\
  ngEHTp1 & EHT(2022)+ & OVRO & HAY & CNI & BAJA & LAS \\
  ngEHT & ngEHTp1+ & GARS & GAM & CAT & BOL & BRZ \\
\end{tabular}
\caption{Array configurations used for EHT and ngEHT coverage and simulated data sets. ngEHT configurations assume participation of existing EHT(2022) sites, as well as the addition of existing/repurposed dishes at HAY (37m), OVRO (10m), and GAM (15m), or hypothetical 6m dishes at new site locations. On-sky bandwidth is assumed to be 4 GHz for EHT(2017), 8 GHz for EHT(2022), and 16 GHz at both 230 and 345 GHz for ngEHT.}
\label{tab:arrayconfiguration}
\end{table}

Before generating the synthetic data, we scatter the movie frames using the interstellar scattering model for Sgr A* by \citet{Johnson2018}, as implemented in \texttt{eht-imaging}. Figure \ref{fig:uv-coverage} presents the uv-coverage of above site arrays. Top row presents the EHT2017(left panel)
and EHT2022(right panel) uv-coverage, while the bottom row shows the uv-coverage for ngEHTp1(left panel) and ngEHT(right panel), respectively. 

\section{Dynamical reconstruction using the StarWarps code} \label{star-warps-code}
Since the gravitational time-scale is substantially short for SgrA* BH, $t_g = GM/c^3 \simeq$ 20 sec, the source structure varies a lot throughout the course of observation. Consequently, the static image assumption \cite{2022ApJ...930L..15A} breaks down in this highly variable source. {\tt StarWarps}  is a novel algorithm which was provided by \citep{2017arXiv171101357B} to model the Very Long Baseline Interferometry (VLBI) observations from a Gaussian Markov Model. StarWarps simultaneously reconstructs both of the image and its motion and it thus allows for an evolving emission region. Likewise the static image reconstruction, it uses the earth rotation synthesis to increase the spatial frequencies out of the earth rotation. However, while in static sources, the VLBI measurements correspond to the same image, for the highly variable sources, they no-longer are related to the same image. In more detail, {\tt StarWarps} reconstructs a $N$ dimensional image vector $X = \lbrace x_1, x_2, ..., x_N \rbrace$ instantaneously, where $N$ referring to total duration of the observation which are taken as sparse observational data array $Y = \lbrace y_1, y_2, ..., y_N \rbrace$. {\tt StarWarps} defines a dynamical imaging model, hereafter called $\varphi$, for each of the observed data as:
\begin{align}
\label{starwarps1}
\varphi_{y_t |x_t} &= \mathcal{N}_{y_t}(f_t(x_t), R_t), \\
\label{static2}
\varphi_{x_t} &= \mathcal{N}_{x_1}(\mu_t, \Lambda_t), \\
\label{dynamic}
\varphi_{x_t |x_{t-1}} &= \mathcal{N}_{x_t}(Ax_{t-1},Q),
\end{align}

\begin{figure*}[th!]
\centering
\includegraphics[width=\textwidth] {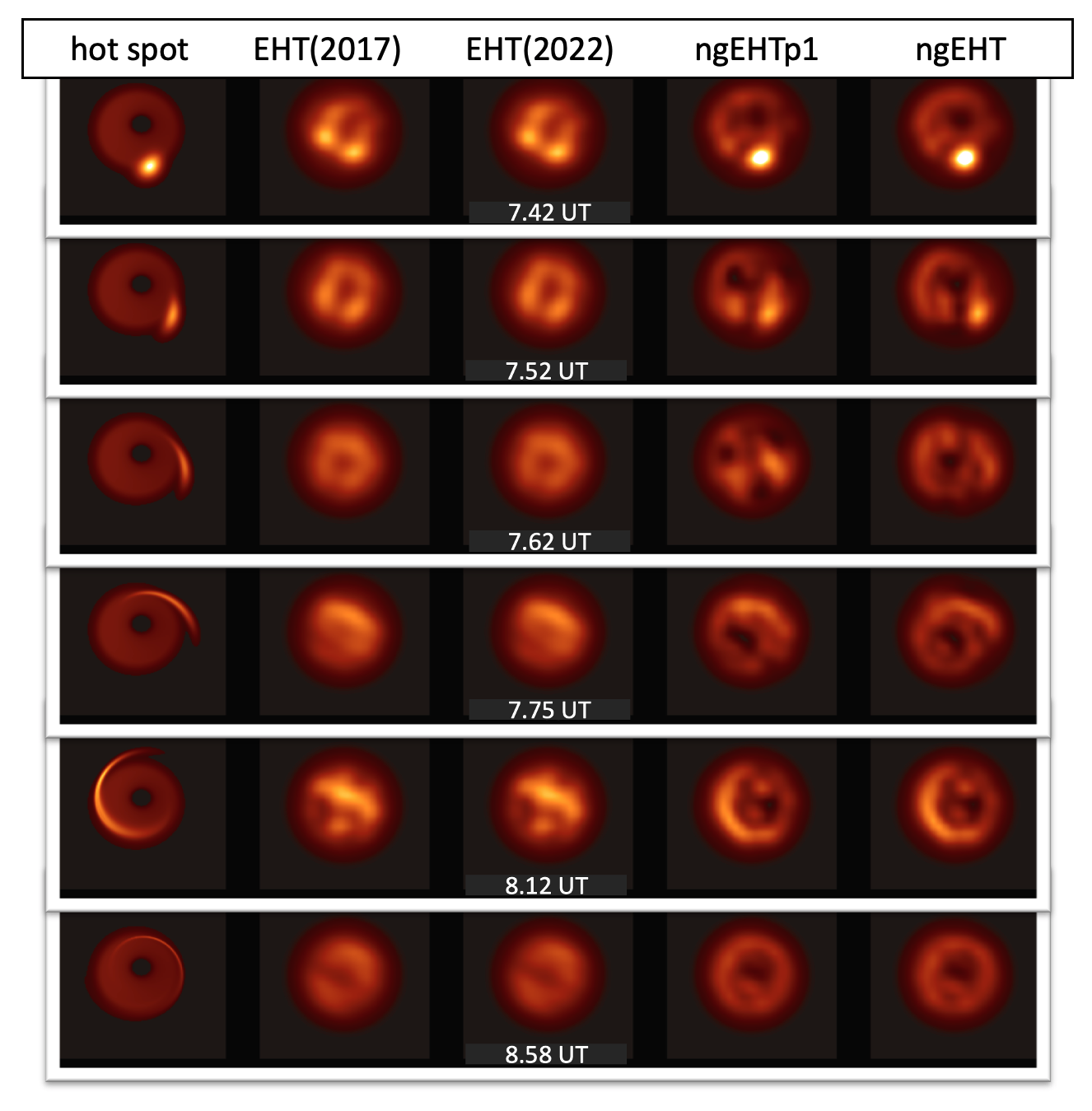}
\caption{Dynamical reconstruction of the orbital motion of the shearing hot spot using few different observational arrays. Down rows correspond to various times. In each row, the 
most left panel refers to the original hot spot model, the second and third columns describe the reconstructed image using the EHT(2017) and EHT(2022) arrays and the last two ones refer to the reconstruction made using ngEHT phase I (ngEHTp1) as well as the full ngEHT array (ngEHT), respectively. It is inferred that the both of the first and full phases of the ngEHT are capable to trace the dynamical motion of the hot spot quite well. }
\label{fig:reconst-hotspot}
\end{figure*}
with $\Lambda_t = diag[\mu_t]^T \Lambda' diag[\mu_t]$. Here, $\mu_t$ describes the mean value of a multivariate Gaussian distribution and $\Lambda$ refers to its covariance. Furthermore, $A$ describes the global time evolution of source image, measuring dynamical evolution of source emission region in a time interval ($t-1$-$t$). Finally, additional variations of the source image is incorporated inside a time invariant covariance matrix $Q$, which highlights the time evolution of the source. Very similar to the static imaging model, each observed $y_t$ corresponds to its analog source image $x_t$, through the function $f_t(x_t)$. However, {\tt StarWarps} also takes into account the dynamical correlation between different snapshots through Eq. \ref{dynamic} with a graceful exit to static imaging when $(A = 1, Q = 0)$. Every image $x_t$ is related to the former image $x_{t-1}$ using $x_t \approx A x_{t-1} $. Every source is treated as a 2D light pulse originated from the angular sky coordinates $(\rho,\delta)$. These pulses leads to some variations in the image by conducting some shifts. 

\begin{figure*}[th!]
\centering
\includegraphics[width=\textwidth]{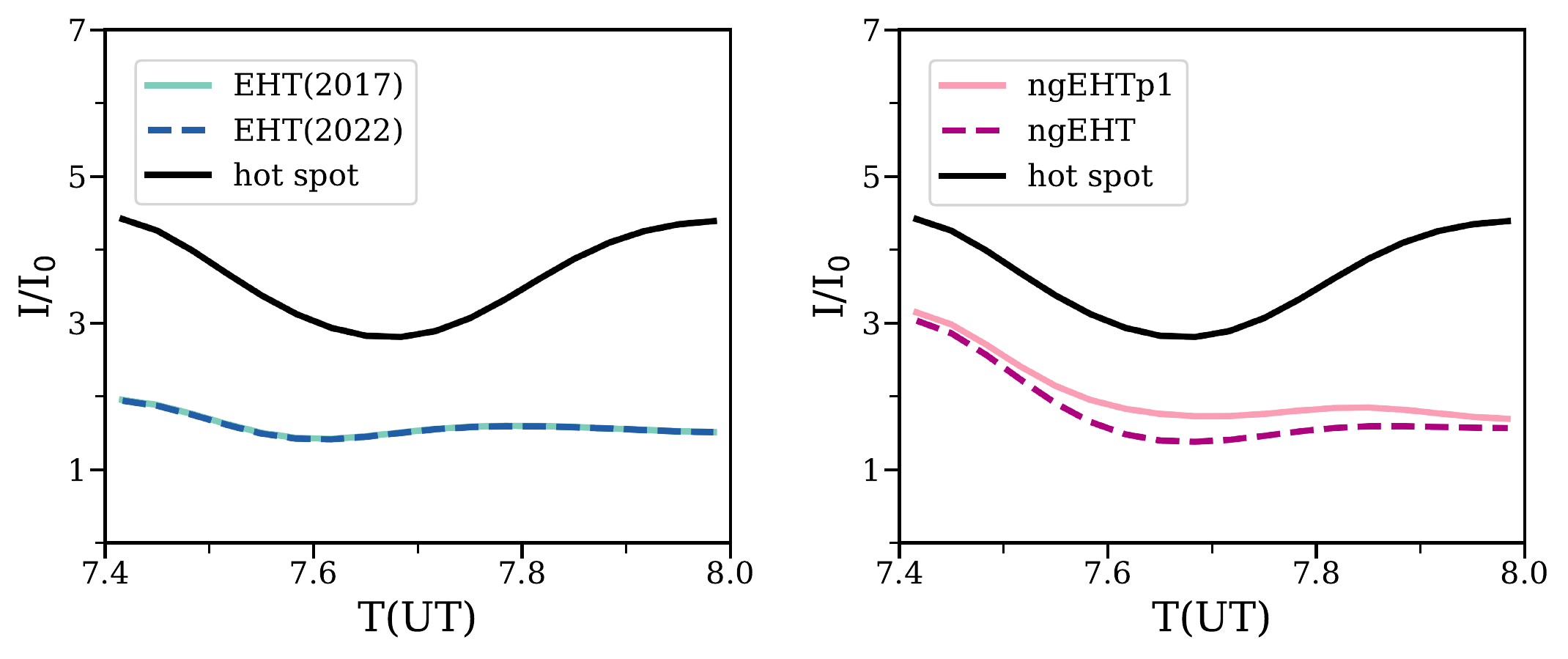}
\includegraphics[width=\textwidth]{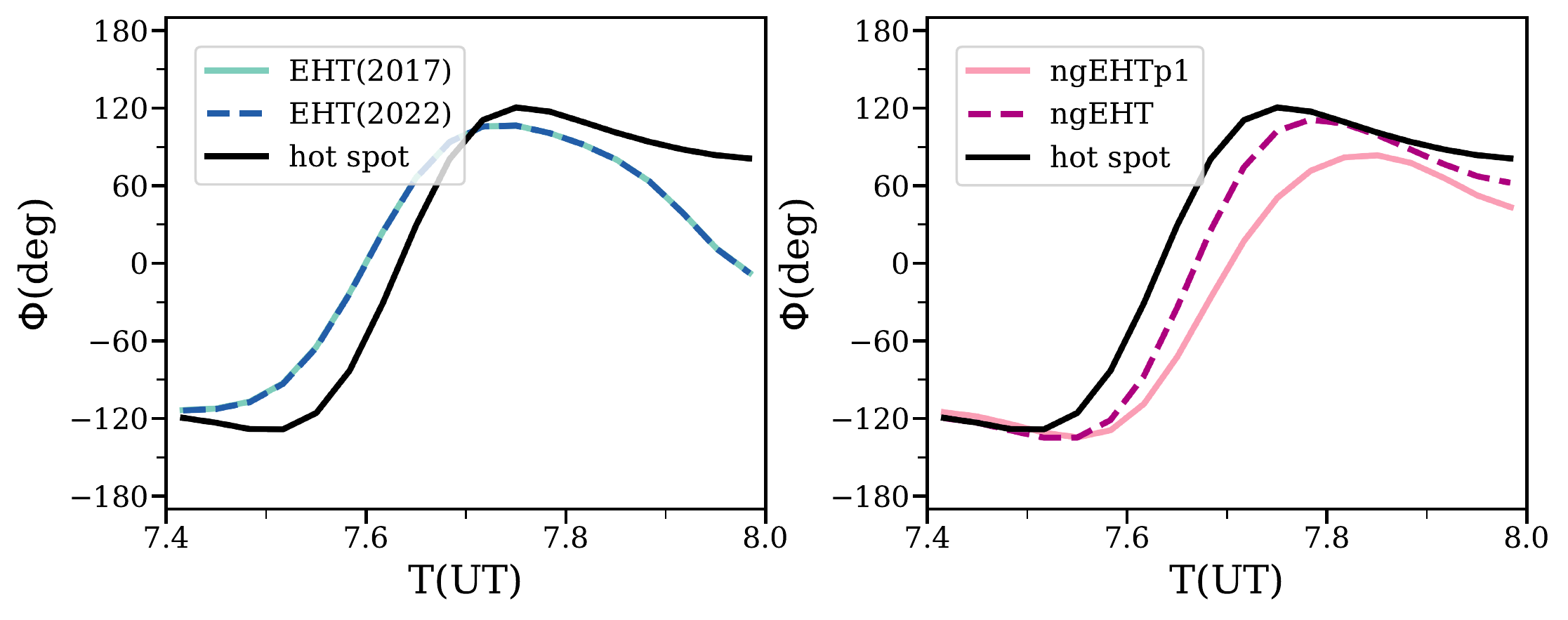}
\caption{The time evolution of the intensity-ratio (top) and phase of the bright-spot (bottom) of the hot spot from the first phase. In each row, the left(right) panel presents EHT(ngEHT) arrays. Overlaid on the plot, we also present the corresponded parameters in the original hot spot. 
It is seen that ngEHT arrays are doing relatively better job in reconstructing the orbital parameters of the hot spot in the first phase. }
\label{fig:firstphsse}
\end{figure*}
{\tt StarWarps} solves for N-D image array $X$, by using N-D observed data points. I this method, $f_t, \mu_t, R_t, \Lambda_t$ and $Q$ are specified parameters. On the other hand, $A$ may or may not be necessarily known. In the latter case, we solve for $A$ jointly with N-D images by making use of the Expectation-Maximization (EM) algorithm. Using this method, we first compute $A$ and then use it to infer $X$.

Finally, the Joint distribution of this dynamical imaging algorithm is computed as:
\begin{equation}
p(X,Y ;A) \propto \prod_{t=1}^N \varphi_{y_t | x_t} \prod_{t=1}^{N} \varphi_{x_t} \prod _{t=2}^{N} \varphi_{x_t | x_{t-1}}. 
\end{equation}
In our reconstructions, we used the bispectrum, visibility amplitude and the log closure amplitude as the data-terms. 2\% systematic noise is added on the top of this and the
a ring with the typical diameter of SgrA* with 25 $\mu$as added blurring is chosen as the prior. Table \ref{chi2-values} presents the $\chi^2$ of different arrays.

\begin{table}
\centering
\caption{Down rows present the data terms as well as different $\chi^2$s for the quality of the reconstructed images using StarWarps for different arrays. From top to bottom we present EHT(2017), EHT(2022), ngEHTp1 and the full ngEHT.  Bias, logcam and cphase refer to the bisepctrum, log closure amplitude and the closure phase, respectively. }
\makebox[ 0.83 \textwidth][c]{
\begin{tabular}{c|cccccc}
\toprule
\hline
Obs  & Bias  & logcam & cphase  & $\chi_{\mathrm{cphase}}^2$ & $\chi_{\mathrm{logcamp}}^2$ & $\chi_{\mathrm{camp}}^2$ \\ \hline
EHT(2017) & 1.0 & 1.0 & 1.0 & 0.67 & 1.16 & 1.37\\ \hline
EHT(2022) & 1.0 & 1.4 & 1.7 & 0.59 & 0.63 & 0.77 \\ \hline
ngEHTp1 & 1.2 & 1.5 & 1.5 & 1.14 & 1.5 & 1.84 \\ \hline
ngEHT & 1.0 & 1.0 & 1.0 & 1.17 & 1.51 & 1.90 \\ \hline
\bottomrule 
\end{tabular}
}
\label{chi2-values}
\end{table}


\section{Reconstructing the motion of hot spot in different arrays}
\label{hotspot-recons}
Here we use the StarWarps code to make a dynamical image reconstruction of the orbiting hot spot using different observational arrays. We subdivide the hot spot motion into two distinct phases, making a novel feature extraction algorithm to trace the orbital motion in both of these phases, respectively. 

\subsection{Tracking the angular location of hot spot} \label{trace-hotspot}
The motion of a shearing hot spot can be subdivided to two distinct phases. The first phase is corresponded to a bright (condensed) blob that initially appears and starts moving around. This motion is then followed by a subsequent phase accorded with the expansion of the hot spot while shearing. To trace both of these features, in the following, we use two distinct metrics. 

$\bullet$ \textbf{\textit{First phase:}} Since the initial blob is condensed, we track its angular location by following the intensity maximum. However, owing to the subsequent shearing of the hot spot, this approximation breaks down soon after the second phase begins.

Figure \ref{fig:firstphsse} compares the time evolution of the normalized intensity (top row) as well as the angular location of the intensity maximum, referred as $\Phi$, (bottom row) between the original hot spot (black-solid-line) and the reconstructed values from different observational arrays. This includes EHT2017(cyan-solid-line) and EHT2022(dashed-blue-line) arrays (left panel) as well as the ngEHTp1(pink-solid-line) and ngEHT(dashed-magenta-line) arrays (right panel), respectively. 

I$_0$ refers to the initial intensity of either the original hot spot or different observational arrays. As the initial flux differs between the original and the reconstructed hot spot, in each case, we normalize the flux to its initial value. 

To make the figure, we have used a gaussian smoothing. Furthermore, as the majority of the data in the second phase can not be described by a condensed bright-spot, we have removed these data. More explicitly, we cut the movie at T $\geq$ 7.65 UT, corresponded to the transition from the first phase to the second one. 

From the plot it is inferred that the reconstructed shape of the intensity and the phase are closer to the original hot spot for ngEHT arrays than the EHT ones. Furthermore, in the phase plot, the bottom row, it is seen that in original and final times the phase is very close to the original hot spot, again the level of agreement is higher in ngEHT arrays than the EHT ones. 

$\bullet$ \textbf{\textit{Second phase:}} This starts when the moving hot spot starts shearing around. Below, we model this motion with an stretched ellipse and infer its axes ratio as well as the ellipse area with time. Since the background is dominated by the RIAF model, to extract the ellipsoidal motion, in each snapshot, we first find out the points with an intensity above 80\% of the intensity max on that snapshot. We then compute the ellipticity as $(\mathrm{b/a})$ where $a$ and $b$ are the associated semi-major and semi-minor ellipse axis, respectively. Furthermore, the ellipse area is also estimated as $\pi a b$. \footnote{we have used SVD from linear algebra in python package.}

\begin{figure*}[th!]
\centering
\includegraphics[width=\textwidth]{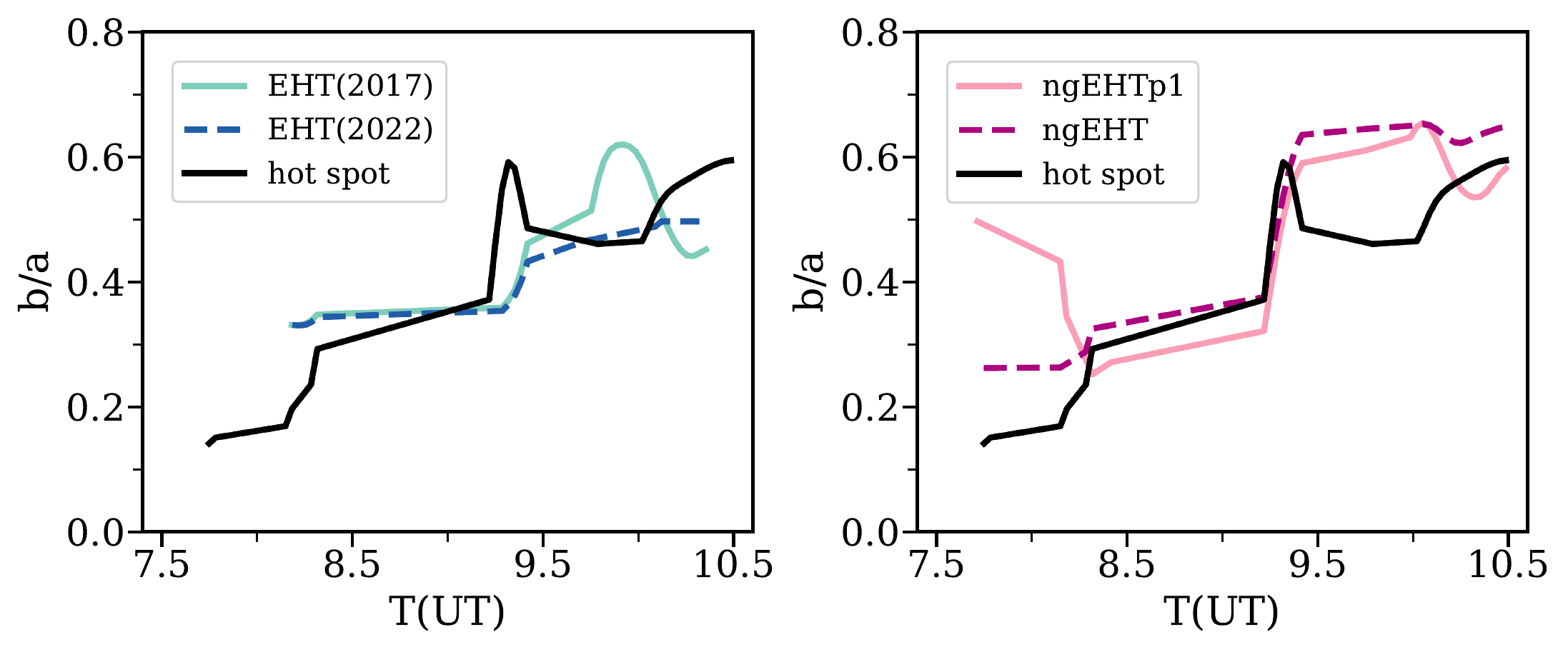}
\includegraphics[width=\textwidth]{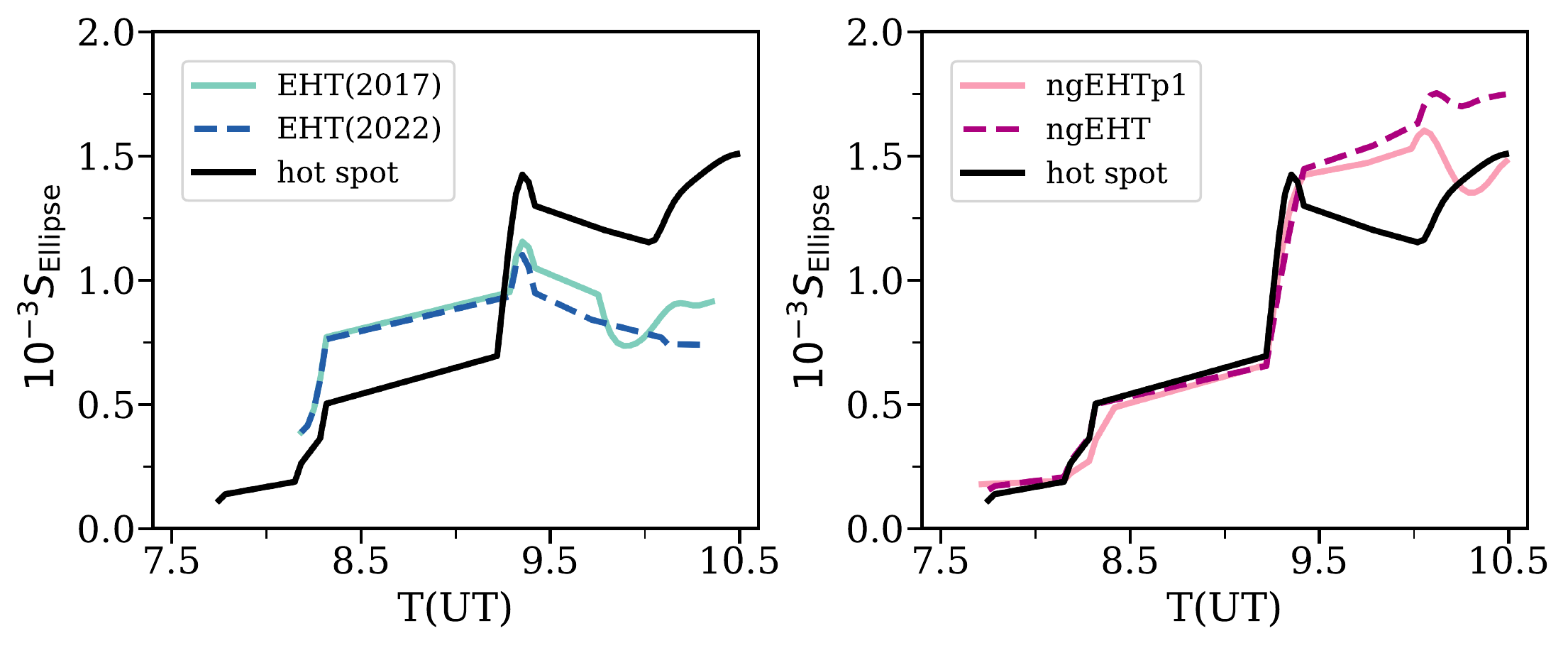}
\caption{The time evolution of the ellipticity (top row) and the ellipse area (bottom row) of the shearing hot spot from the second phase. Overlaid on the plot, we also present the original hot spot, different EHT (left) and ngEHT (right) arrays. It is seen that up until the time that hot spot decays its motion significantly, ngEHT does a relatively good job in recovering the actual motion.}
\label{fig:Second-Phase}
\end{figure*}

Figure \ref{fig:Second-Phase} presents the time evolution of the ellipticity (top row) and the ellipse area (bottom row) using different observational arrays. Overlaid on the plot, we also show the corresponding values for the hot spot model. It is clearly seen that the ngEHT arrays work better in reconstructing the elliptical motion. To make the plot more readable, we only show the snapshots for them the area of the reconstructed image, using individual arrays, sits between 0.5-2.0 of the original hot spot. This removes some of the snapshots where the reconstruction is not too ideal. Furthermore, the ellipse area is fairly similar between the original hot spot and different ngEHT phases. The EHT associated ellipse area, on the contrary, establish less similarity with the original hot spot. 

\begin{figure*}[th!]
\centering
\includegraphics[width=\textwidth]{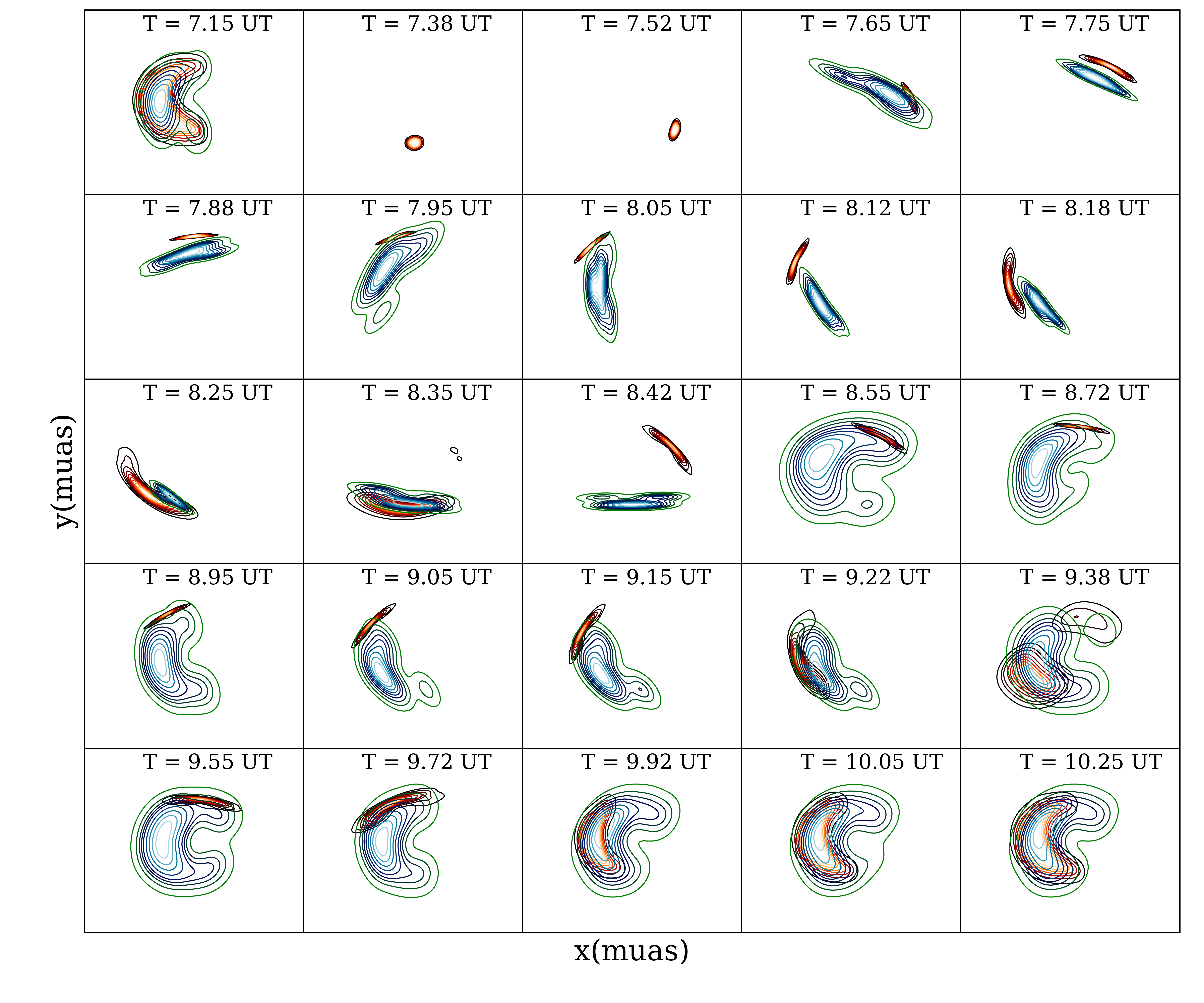}
\caption{Comparison between the extracted trajectory of the original hot spot (red color map) and from the ngEHT reconstructed image (blue color map) at 25 different snapshot. It is seen that overall ngEHT can track the motion of the hot spot quite good. It's performance is however much better in some snapshots than the others. The plot shows the KDE of points with an intensity above 80\% of the intensity max. }
\label{fig:Ellipse-Motion}
\end{figure*}

Figure \ref{fig:Ellipse-Motion} presents the extracted elliptical motion for the original hot spot (red color map) as well as the ngEHT (blue color map) at few different snapshots. To make the plot more readable, we skip showing the trajectory for the EHT arrays and the ngEHTp1 array. From the plot, it is clearly seen that some snapshots do a really great job in reconstructing the actual motion of the hot spot, others may however overestimate the area of the ellipse. This is not surprising as we are limited by the resolution. This somewhat explains why we needed to remove them in an unbiased comparison between the ellipticity and the area of the ellipse. 
\subsection{$\mathrm{Nxcorr}$ vs $\mathrm{Nrmse}$ of the reconstructed and ground truth image}
To make the comparison between the reconstructed and the ground truth images more quantitative, here we compute the normalized cross-correlation (hereafter $\mathrm{Nxcorr}$) as well as the normalized root-mean-squared error (hereafter $\mathrm{Nrmse}$) between the reconstructed image and its ground truth image.

$\bullet$ $\mathrm{Nxcorr}$: 
We make use of  \cite{2019ApJ...875L...4E} and \cite{2018zndo...1173414C} defining the $\mathrm{Nxcorr}$ as:
\begin{equation}
\label{NXcor}
\mathrm{Nxcorr}(X,Y) = \frac{1}{N} \sum_{i} \frac{\left(X_i - \langle X \rangle \right)\left(Y_i - \langle Y \rangle \right)}{\sigma_X \sigma_Y},
\end{equation}
where $X$ refers to the restructured image, while $Y$ describes the ground truth image of the hot spot. Furthermore, $N$ stands for the umber of the pixels in the image and $\langle  \rangle$ refers to the mean pixel value of the image. Finally, $\sigma_i$ describes the standard deviation of pixel values in image $i$. 
$\mathrm{Nxcorr}$ determines the similarities between two images. A perfect correlation between the images leads to 1, while a complete anti-correlation between them gives rise to a value of -1  for $\mathrm{Nxcorr}$. 

$\bullet$ $\mathrm{Nrmse}$: is defined as \cite{2018zndo...1173414C}:
\begin{equation}
\label{Nrmse}
\mathrm{Nrmse} = \frac{\sum_{i} |X_i - Y_i|^2}{\sum_{i} |X_i|^2}.
\end{equation}
where ulike the case of $\mathrm{Nxcorr}$, two completely similar(different) images $X$ and $Y$ have 0 (1) value $\mathrm{Nrmse}$.

\begin{figure*}
\centering
\includegraphics[width=\textwidth]{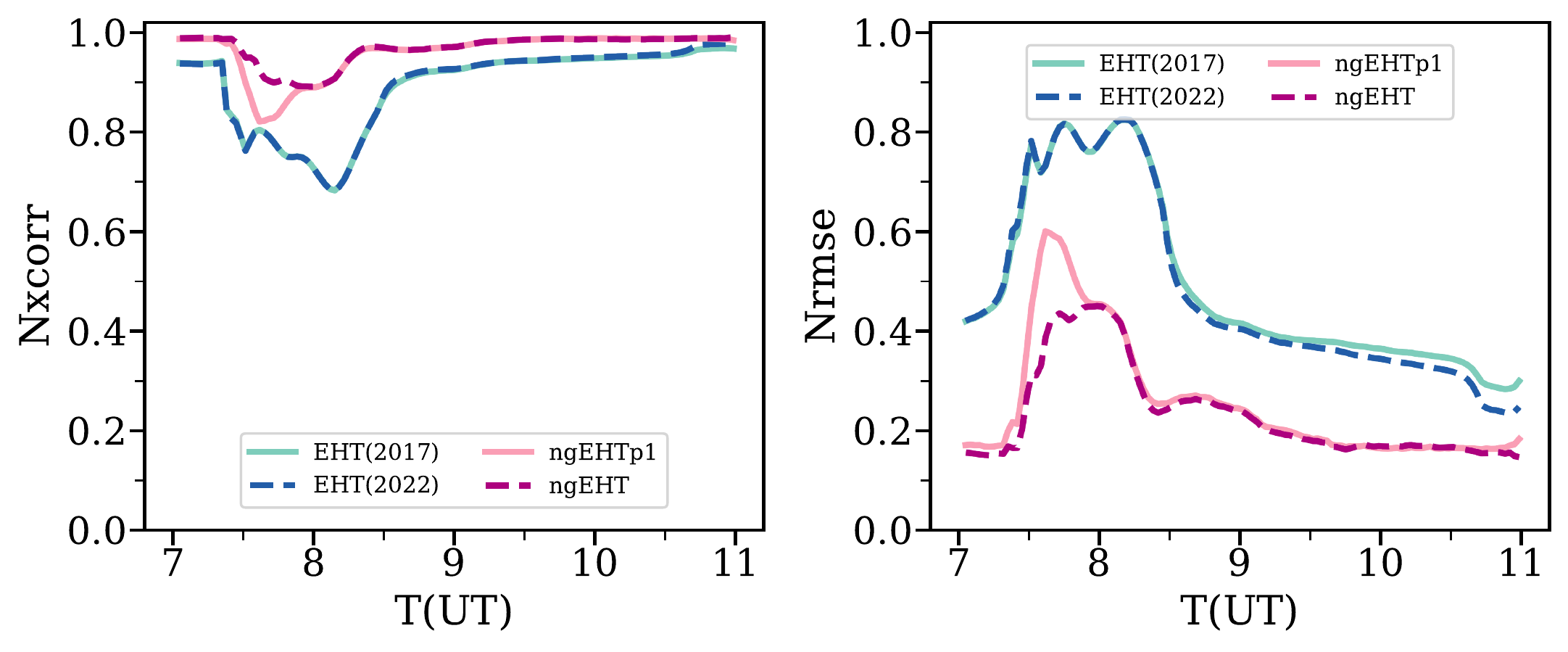}
\caption{$\mathrm{Nxcorr}$ (left panel) and the $\mathrm{Nrmse}$ (right panel) for reconstructed shearing hot spot using different observational arrays. During the shearing phase of the hot spot, there is some suppression(enhancement) of $\mathrm{Nxcorr}$($\mathrm{Nrmse}$) from the pure background RIAF. The deviation is however minimal for ngEHT array compared with both of EHT(2017) and EHT(2022) arrays.}
\label{fig:correlation}
\end{figure*}

Figure \ref{fig:correlation} presents the $\mathrm{Nxcorr}$
and $\mathrm{Nrmse}$ for reconstructed images computed using different arrays. From the plot, it is inferred that: 

$\circ$ Since the background RIAF is dominated in some snapshots, it is seen that we have a globally good correlation between the images. 

$\circ$ This is however getting worse when the hot spot appears and get sheared down, in which it is seen that we have a some levels of suppression(enhancement) of $\mathrm{Nxcorr}$($\mathrm{Nrmse}$) for some cases. 

$\circ$ The aforementioned suppression(enhancement) is however minimal for the ngEHT array compared with the EHT(2017) and EHT(2022). 

$\circ$ Consequently, we conclude that ngEHT array helps a lot in improving the quality of the reconstructed image. 

\section{Conclusion}
We made an in-depth study of tracing the dynamical motion of a shearing hot spot, proposed in \cite{2020ApJ...892..132T}, using StarWarps package \cite{2017arXiv171101357B}, a dynamical image reconstruction algorithm, employing different observational arrays, including both of the EHT as well as the ngEHT arrays. We subdivided the dynamical orbital motion of the hot spot to two distinct phases, that are also observed in GRMHD simulations (see Figure \ref{fig:Plasmoid-hotspot}),  and traced the motion in each of these phases, respectively. The first phase focuses on the appearance of the hot spot and its initial motion when it is ejected from the reconnection layer, while the second phase explores the shearing of the hot spot (potentially driven by Rayleigh-Taylor instabilities at the hot spot boundary during its orbit), being modeled with a re-shaping ellipse. Leptons originating from the jet, accelerated through an equatorial reconnection layer, may end up in the orbiting hot spot confined by vertical magnetic field. They can then go through a secondary acceleration phase due to the shearing motion. It is conjectured in \citep{2022ApJ...924L..32R} that such accelerated leptons in the hot spot can power NIR flares and potentially concurrent submm emission.
We made a novel algorithm to trace the orbital phase in the first phase and the axes ratio and the ellipse area in the second phase. Furthermore, we inferred the Nxcorr and the Nrmse for different observational arrays. Our analysis showed that while EHT arrays might have some difficulties in locating the hot spot in the first phase, which gets even harder to trace the motion in the shearing phase, adding more sites to the array, as is planned in the ngEHT, substantially helps to improve the quality of the reconstructed image in both phases. Consequently, we propose to use the ngEHT to trace the dynamical motion of the hot spot. While the analysis done in this work is only limited to the hot spot, we argue that the dynamical reconstruction and feature extraction algorithms used in this study can be easily extended to any types of dynamical motions.

In this work, we only addressed the issues related to the total intensity modeling of hot spots that could be observed with the ngEHT. However, hot spots emerging in the accretion flow may indicate significant fractional linear polarization. Since mm wavelength radiation in Sgr~A* originates through the synchrotron process, this allows to probe the magnetic field geometry with hot spots through imaging of the linear polarization, e.g., \citep{Vos2022}. Indeed, the linear polarization observations of unresolved Sgr~A* provided a strong argument for orbiting hot spot model of flares \citep{gravity_loops_2018,Wielgus2022b}. While more comprehensive studies are necessary to address this subject, it is clear that resolving the polarized structure of the source with the ngEHT will vastly improve our understanding of the magnetic field geometry and time-evolution.

\section{Acknowledgements}

It is a great pleasure to acknowledge Katherine L. Bouman and Michael Johnson for very fruitful conversations. Razieh Emami acknowledges the support by the Institute for Theory and Computation at the Center for Astrophysics as well as grant numbers 21-atp21-0077, NSF AST-1816420 and HST-GO-16173.001-A for very generous supports. We thank the supercomputer facility at Harvard where most of the simulation work was done. FR was supported by NSF grants AST-1935980 and AST-203430. AR and KC acknowledges support by the National Science Foundation under Grant No. OISE 1743747 as well as the support of grants from the Gordon and Betty Moore Foundation and the John Templeton Foundation. PN gratefully acknowledges support at the Black Hole Initiative (BHI) at Harvard as an external PI with grants from the Gordon and Betty Moore Foundation and the John Templeton Foundation.

\section{References}
\bibliography{ms}
 
\end{document}